\documentclass[11pt,a4paper]{article}
\usepackage[margin=23mm]{geometry}
\usepackage{amsmath,amssymb,mathtools,bm}
\usepackage{graphicx,booktabs,array,microtype}
\usepackage{enumitem}
\usepackage{placeins}
\usepackage[hidelinks]{hyperref}
\usepackage{fancyhdr}
\allowdisplaybreaks

\newcommand{\R}{\mathbb R}
\newcommand{\N}{\mathcal N}
\newcommand{\W}{\mathcal W}
\newcommand{\dd}{\mathrm d}
\newcommand{\norm}[1]{\left\lVert #1\right\rVert}
\newcommand{\ip}[2]{\left\langle #1,#2\right\rangle}

\newcommand{\diag}{\operatorname{diag}}

\title{\bfseries Spectral rigidity beyond Braak's conjecture\\
in the quantum Rabi model}
\author{Murray T. Batchelor$^{1,2}$ \quad and \quad Zi-Min Li$^{3}$\\[6pt]
\small $^1$Mathematical Sciences Institute, Australian National University,\\
\small Canberra ACT 2601, Australia\\[4pt]
\small $^2$Centre for Modern Physics, Chongqing University,\\
\small Chongqing 400044, The People's Republic of China\\[4pt]
\small $^3$Institute of Quantum Physics, Hunan Key Laboratory of Nanophotonics and Devices,\\
\small School of Physics, Central South University, Changsha 410083, China}
\hypersetup{
 pdftitle={Spectral rigidity beyond Braak's conjecture in the quantum Rabi model},
 pdfauthor={Murray T. Batchelor and Zi-Min Li}
}

\begin{document}
\maketitle

\begin{abstract}
Braak's exact solution of the quantum Rabi model determines the regular spectrum through the zeros of spectral $G$-functions. His additional spectral-counting conjecture places bounds on the number of levels of each parity between successive integer pole positions. We show that these interval counts admit no finite upper bound uniform in the coupling and atomic splitting. In a joint limit where the atomic splitting grows as the square of the coupling, every fixed interval between successive nonnegative integers in shifted energy contains logarithmically many levels of each parity. The accumulation occurs near the separatrix of the lower adiabatic potential. We obtain the corresponding parity-resolved count for the full quantum operator with a uniformly bounded remainder. Two exact constraints survive. The cumulative parity counts differ by at most one at every energy, and the conjectured counting rules hold above an explicit threshold determined by the coupling and atomic splitting. The macroscopic density within each parity sector equals the oscillator value above the upper-band onset. A universal quartic crossover governs counts growing as the square root of the coupling in windows at negative shifted energies.
\end{abstract}

\section{Introduction}\label{sec:intro}

The quantum Rabi model describes a two-level system coupled
to a single quantized bosonic mode. Its semiclassical
precursor goes back to Rabi's studies of spin dynamics in
time-dependent magnetic fields~\cite{Rabi1936,Rabi1937}. The eightieth
anniversary of Rabi's original work was commemorated in
Ref.~\cite{Braak2016}. Reviews of the exact solution, dynamics and
light-matter applications are given in Refs.~\cite{Xie2017,FornDiaz2019,Kockum2019,LeBoite2020,DeBernardis2024}.
In units $\hbar=\omega=1$, the Hamiltonian is
\begin{equation}
 H=a^\dagger a+g(a+a^\dagger)\sigma_x+\Delta\sigma_z,
 \qquad g>0,
 \label{eq:H}
\end{equation}
where $2|\Delta|$ is the bare atomic level splitting. The parity operator is $\Pi=(-1)^{a^\dagger a}\sigma_z$, and the spectrum is simple within each parity sector. 
Let $E_j^s$, with $s=\pm$ and $j=0,1,2,\ldots$, denote the ordered eigenvalues in each parity sector. We introduce the shifted energies
\begin{equation}
 x_j^s=E_j^s+g^2.
 \label{eq:shift}
\end{equation}

Braak's exact solution~\cite{Braak2011} gives the regular spectrum through the zeros of explicitly defined functions $G_s(x)$, one for each parity. Chen et al. recovered these functions using Bogoliubov operators and extended coherent states~\cite{Chen2012}. Zhong et al. obtained analytical eigenstates in terms of confluent Heun functions, encompassing regular and exceptional solutions~\cite{Zhong2013}. Braak also introduced a generalized $G$-function to improve numerical access to the high-energy spectrum~\cite{Braak2013}. 
The possible poles of the spectral \(G\)-functions occur at the nonnegative integers in shifted energy. Braak’s spectral-counting conjecture concerns the number of energy levels of each parity between these successive integer points.

For each parity, the number of shifted eigenvalues in the half-open unit interval $[n,n+1)$ is
\begin{equation}
 B_n^s=\#\{j:n\le x_j^s<n+1\},
 \qquad n\in\mathbb Z_{\ge0}.
 \label{eq:B}
\end{equation}
We count the full parity spectrum, including exceptional eigenvalues. Our half-open convention assigns an eigenvalue at an integer endpoint to the interval beginning at that point. When neither endpoint is an eigenvalue, this agrees with counting regular $G_s$-function zeros between successive poles.

Braak's spectral-counting conjecture, also called the $G$-conjecture~\cite{Braak2019}, imposes the counting inequalities 
\begin{equation}
 B_n^s\le2,
 \qquad 1\le B_n^s+B_{n+1}^s\le3.
 \label{eq:Braak}
\end{equation}
Our convention extends these inequalities to exceptional parameter values by including integer eigenvalues as specified above. These counting rules are additional to the exact quantization condition. The recent counterexamples reported by Lanuza \cite{Lanuza2026} show that the rules can fail without affecting the validity of the $G$-function solution. Here we ask how large the violations can become and what exact constraints remain. 
We use spectral rigidity to mean exact restrictions on the counting and relative ordering of the two parity spectra.

Our main result is that no finite parameter-independent bound on these counts can replace the conjectured cap. For each fixed interval $[n,n+1)$ with $n\ge0$ and either parity,
\begin{equation}
 \Delta=g^2-\frac12,
 \qquad B_n^s=\frac{\log g}{\pi}+O_n(1),
 \qquad g\longrightarrow\infty.
 \label{eq:introfixedbin}
\end{equation}
Thus even the fixed interval $[0,1)$ contains arbitrarily many levels of each parity. We refer to this increasing concentration of levels within an interval of fixed width as spectral crowding. The same parameter path gives this growth in every fixed interval $[n,n+1)$ with $n\ge0$, with a remainder uniform over any prescribed finite collection of such intervals.

We establish that the accumulation occurs near the separatrix of the lower adiabatic double-well potential, where classical motion slows near the barrier. The associated logarithmic singularity of the density of states was identified in studies of the excited-state quantum phase transition \cite{Puebla}. 
Related studies of quantum scrambling in anisotropic Rabi and Dicke models connect spectral singularities to the geometry and local stability of semiclassical stationary points~\cite{Li2026Scrambling,Li2026ChaosScrambling}.
Here we derive a parity-resolved counting law for the full quantum operator, with a bounded remainder uniform in compact barrier-relative energy windows, including windows whose endpoints move with the coupling.

Two exact counting constraints survive. At every energy and for arbitrary parameters, the strict cumulative counts satisfy
\begin{equation}
 \mathcal N_s(E)=\#\{j:E_j^s<E\},
 \qquad |\mathcal N_+(E)-\mathcal N_-(E)|\le1.
 \label{eq:introparity}
\end{equation}
Braak's counting rules are also restored above an explicit sufficient threshold,
\begin{equation}
 n\ge g^2+|\Delta|
 \quad\Longrightarrow\quad
 B_n^s\le2,\qquad 1\le B_n^s+B_{n+1}^s\le3.
 \label{eq:introtail}
\end{equation}
The first relation follows from the half-line boundary conditions of the two parity sectors. The second follows by excluding non-Juddian pole eigenvalues above the threshold and identifying the ordered index of every Juddian crossing. Arbitrarily strong local crowding therefore coexists with bounded parity imbalance and an exactly constrained tail.

The adiabatic description places these results in the full spectrum. Above the onset of the upper band, the macroscopic same-parity density equals the oscillator value. At the endpoint where the lower barrier becomes quartic, a universal crossover governs local counts of order $\sqrt g$. These windows lie at negative shifted energies and describe a separate regime from the unit intervals at nonnegative shifted energies in the conjecture.

Large-index asymptotics at fixed coupling and atomic splitting were established by Boutet de Monvel and Zielinski \cite{BdMZ}. Using these asymptotics, Rudnick proved a density-one version of Braak's conjecture \cite[Theorem~1.2]{Rudnick}. At fixed positive parameters, a set of integer intervals of natural density one has two shifted levels of one parity and none of the other, with the parity assignments reversed in the neighbouring intervals. Our tail theorem establishes the counting inequalities in every interval above an explicit sufficient threshold. It therefore excludes violations even in the sparse set not controlled by the asymptotic argument.

The orders of limits are essential. Finite-window validity at sufficiently small atomic gap was proved by Malagutti and Parmeggiani \cite{MPFerrara,MP2026}. Our crowding limit increases the coupling and atomic splitting together. The interval in shifted energy can remain fixed while its location in physical energy and the ordered indices of its levels change. This is compatible with exact tail restoration at every fixed parameter pair.

In establishing the above results we begin with the parity spectra and their counting relation. We then derive the unbounded separatrix count and establish restoration of the counting rules. The macroscopic density and quartic crossover complete the spectral picture. Detailed working is left to the Appendices.

\section{Parity spectra and the exact counting relation}\label{sec:parity}
For completeness, the spectral functions introduced above are
\begin{equation}
 G_s(x)=\sum_{m=0}^{\infty}K_m(x)
 \left(1-\frac{s\Delta}{x-m}\right)g^m,
 \qquad s=\pm1.
 \label{eq:Gfunction}
\end{equation}
The dependence of $K_m$ on $g$ and $\Delta$ is suppressed. The coefficients are fixed by
\begin{equation}
 K_0=1,\qquad K_1(x)=f_0(x),\qquad
 mK_m(x)=f_{m-1}(x)K_{m-1}(x)-K_{m-2}(x),\quad m\ge2,
 \label{eq:Grecurrence}
\end{equation}
where
\begin{equation}
 f_m(x)=2g+\frac{1}{2g}
 \left(m-x+\frac{\Delta^2}{x-m}\right).
 \label{eq:Gcoefficient}
\end{equation}
Their regular zeros give the parity eigenvalues away from nonnegative integer $x$. The exceptional eigenvalues at the integer points are considered in Section~\ref{sec:tail}.

In the bosonic number basis the parity sectors are the Jacobi matrices
\begin{equation}
 (J_s)_{mm}=m+s\Delta(-1)^m,\qquad
 (J_s)_{m,m+1}=g\sqrt{m+1},\qquad m\ge0.
 \label{eq:jacobi}
\end{equation}
For $g>0$, the eigenvalue equation determines every component of an eigenvector from its first component, which cannot vanish for a nonzero eigenvector. Thus each parity eigenvalue is simple.
To compare them, write \eqref{eq:H} in the $\sigma_z$ basis on the coordinate half-line $q>0$,
\begin{equation}
 H=-\frac12\partial_q^2+\left(\frac{q^2}{2}-\frac12\right)I
       +\sqrt2gq\,\sigma_x+\Delta\sigma_z.
 \label{eq:coordinateH}
\end{equation}
Positive parity imposes Neumann--Dirichlet conditions on the two spin components, while negative parity imposes Dirichlet--Neumann conditions. The linear coupling is infinitesimally form bounded relative to the oscillator, so these realizations are bounded below and have compact resolvent. 

We use the min--max principle, the variational characterization of ordered energy levels. 
With the lowest level labelled $j=0$, the $j$th eigenvalue is obtained by minimizing, over all $(j+1)$-dimensional trial spaces in the form domain, the largest energy expectation among normalized states in each space.
Restricting the allowed states therefore cannot lower an eigenvalue. Imposing one independent linear constraint gives interlacing, with each restricted eigenvalue lying between two consecutive unrestricted eigenvalues.

Let $d_j$ be the spectrum obtained by imposing Dirichlet conditions on both components. The resulting form domain has codimension one in either parity form domain, since it adds one independent boundary trace condition. Applying the min--max principle gives
\begin{equation}
 E_j^\pm\le d_j\le E_{j+1}^\pm.
\end{equation}
Consequently
\begin{equation}
 E_j^+\le E_{j+1}^-,\qquad E_j^-\le E_{j+1}^+,
 \label{eq:crossinterlace}
\end{equation}
or, equivalently,
\begin{equation}
 |\N_+(E)-\N_-(E)|\le1.
 \label{eq:countdisc}
\end{equation}
Subtracting the strict counts at the endpoints of any interval gives
\begin{equation}
 \left|\#\{j:a\le E_j^+<b\}-\#\{j:a\le E_j^-<b\}\right|\le2.
 \label{eq:windowdisc}
\end{equation}
This exact relative constraint allows consecutive levels of the same parity and arbitrarily large local level counts. It also ensures that a divergent leading count obtained for one parity holds for the other with at most a bounded change.

\section{Unbounded crowding near the separatrix}\label{sec:crowding}
The local count becomes large near the barrier of the lower adiabatic potential. Classical motion slows there, producing a logarithmic increase in travel time. The exact relation \eqref{eq:windowdisc} already ensures that the two parity counts in a common interval differ by at most two. We now determine their common divergent leading term.

Increasing the coupling at fixed atomic splitting cannot make the number of levels in a unit interval grow without bound. At $\Delta=0$, each parity spectrum is the displaced oscillator ladder $j-g^2$. The perturbation $s\Delta(-1)^m$ in \eqref{eq:jacobi} has norm $|\Delta|$, so it changes the energy expectation of any normalized state by at most $|\Delta|$. The min--max principle therefore gives
\begin{equation}
 |E_j^s-(j-g^2)|\le|\Delta|.
 \label{eq:fixedsplittingbound}
\end{equation}
If $x_j^s\in[n,n+1)$, its index lies in $[n-|\Delta|,n+1+|\Delta|)$. The number of possible indices is therefore bounded independently of $g$ when $\Delta$ is fixed.

Instead, consider the joint scaling
\begin{equation}
 \Delta=\beta g^2-c,\qquad 0<\beta<2,
 \label{eq:betascaling}
\end{equation}
with fixed $c$. Here $\beta$ is the limiting ratio $\Delta/g^2$, while the offset $c$ controls the finite position of the barrier. The atomic term remains comparable to the displacement energy.
Diagonalizing the spin matrix at each fixed oscillator coordinate gives two adiabatic potentials. 
The lower potential is
\begin{equation}
 V_-(q)=\frac{q^2}{2}-\frac12-\sqrt{\Delta^2+2g^2q^2}.
 \label{eq:Vminus}
\end{equation}
For sufficiently large $g$, this is a symmetric double well. The barrier at the origin separates motion confined to either well from motion passing through both wells. Its physical and shifted energies are
\begin{equation}
 e_{\rm sep}=-\Delta-\frac12,\qquad
 x_{\rm sep}=e_{\rm sep}+g^2=(1-\beta)g^2+c-\frac12.
 \label{eq:xsep}
\end{equation}

\begin{figure}[htbp]
\centering
\includegraphics[width=0.78\linewidth]{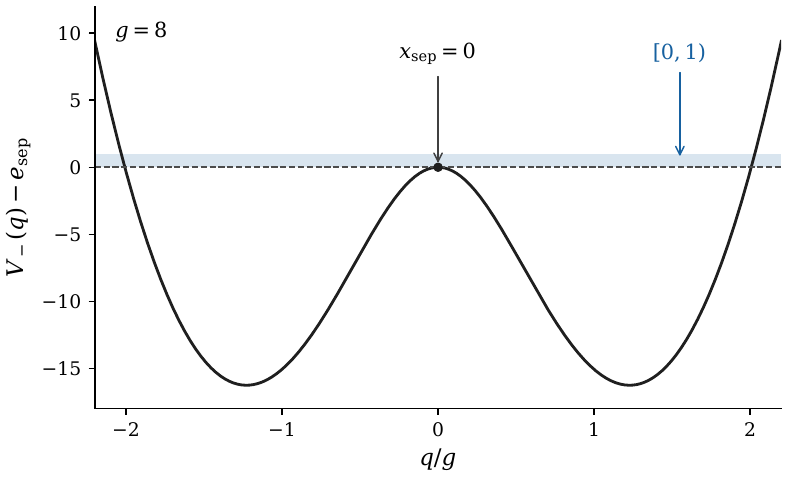}
\caption{The lower adiabatic double-well potential relative to the barrier energy at $g=8$ and $\Delta=g^2-1/2$. The barrier has shifted energy $x_{\rm sep}=0$, and the shaded strip marks the fixed interval $[0,1)$.}
\label{fig:barrier}
\end{figure}

The choice $\beta=1$, $c=1/2$ fixes the barrier at $x_{\rm sep}=0$, as shown in Figure~\ref{fig:barrier}. Along this path, the count in each fixed interval $[n,n+1)$ with $n\ge0$ satisfies
\begin{equation}
 B_n^s\bigl(g,g^2-\tfrac12\bigr)
 =\frac{\log g}{\pi}+O_n(1)\longrightarrow\infty,
 \qquad s=\pm.
 \label{eq:fixedbin}
\end{equation}
The remainder is uniform over any prescribed finite collection of these intervals. Its constant may depend on that set, and no uniform assertion over all $n\ge0$ is required.

\subsection{Finite-coupling illustration}
\begin{figure}[htbp]
\centering
\includegraphics[width=0.78\linewidth]{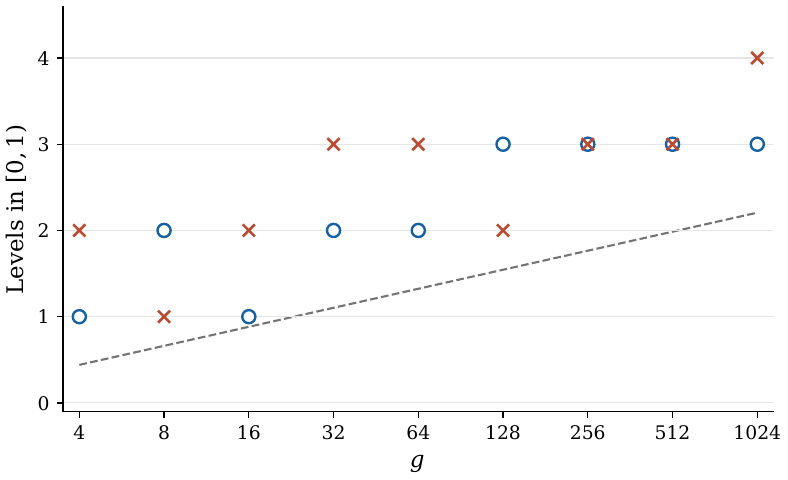}
\caption{Onset of crowding in the fixed interval $[0,1)$ along $\Delta=g^2-1/2$. Blue circles and red crosses denote positive and negative parity, respectively. The counts for each parity at $g=2^k$, $k=2,\ldots,10$, are compared with the leading term $(\log g)/\pi$ without a fitted offset. The $g=1024$ counts are validated for the infinite Jacobi operator. The remaining points agree at two truncation dimensions. Details are given in Appendix~\ref{app:numerics}. The finite examples illustrate the asymptotic law \eqref{eq:fixedbin}.}
\label{fig:fixedbin}
\end{figure}

Figure~\ref{fig:fixedbin} shows the parity-resolved level counts in $[0,1)$ along $\Delta=g^2-1/2$. At $g=32$ and $64$ this interval contains three negative-parity levels. At $g=256$ and $512$ it contains three levels of each parity. At $g=1024$ and $\Delta=1048575.5$, the four negative-parity shifted levels are
\begin{equation}
 x^-\simeq0.004941,\quad0.287491,\quad0.588460,\quad0.905939.
 \label{eq:fourlevel}
\end{equation}
Thus $B_0^-=4$, while $B_0^+=3$ at the same parameters. Outward-rounded Sturm counts and an explicit bound on the omitted oscillator tail validate these counts for the infinite Jacobi operator. The certification method and enclosure widths are described in Appendix~\ref{app:numerics}. The other plotted counts are checked at two finite truncation dimensions.

The integer counts fluctuate and need not be monotone in $g$, consistently with the bounded remainder in \eqref{eq:fixedbin}. The curve $(\log g)/\pi$ shows the leading term without a fitted offset. The analytic counting law establishes unbounded growth independently of these finite examples.

Large parameters require care because a unit interval must be resolved near physical energy $-g^2$. The oscillator cutoffs used here grow as $g^2$, while the number of levels in the interval grows only as $\log g$. This disparity makes large violations demanding to detect in finite parameter scans.

\FloatBarrier

\subsection{The barrier-relative counting law}
The counting law for a fixed interval follows from a more general local count. Near the barrier,
\begin{equation}
 V_-(q)-e_{\rm sep}
 =-\frac{\kappa_\beta^2}{2}q^2+O\!\left(g^{-2}(q^2+q^4)\right),
 \qquad \kappa_\beta=\sqrt{\frac{2-\beta}{\beta}}.
 \label{eq:kappa}
\end{equation}
For a barrier-relative window, write
\begin{equation}
 \W_s^{(\beta)}(a,b;g,c)=\#\{j:a\le x_j^s-x_{\rm sep}<b\}.
 \label{eq:barrierwindow}
\end{equation}
Keep $c\in\R$ fixed, take $a<b$ in a compact interval $\mathcal K\subset\R$, and restrict $\beta$ to a compact set $\mathcal B=[\beta_0,\beta_1]\Subset(0,2)$. Uniformly over these endpoints and parameters, for either parity,
\begin{equation}
 \W_s^{(\beta)}(a,b;g,c)
 =\frac{b-a}{\pi\kappa_\beta}\log g+O(1).
 \label{eq:betacrowd}
\end{equation}
At $\beta=1$, the coefficient is $(b-a)/\pi$. Taking $c=1/2$ and endpoints $n,n+1$ gives \eqref{eq:fixedbin}. More generally, any fixed offset in $\Delta=g^2-c$ leaves the barrier at the finite shifted energy $c-1/2$. The same argument gives $B_n^s(g,g^2-c)=(\log g)/\pi+O_{n,c}(1)$ for every fixed $n\ge0$.

The double-well separatrix exists throughout $0<\beta<2$. For $0<\beta<1$ it moves through positive shifted energies. With $n_g=\lfloor x_{\rm sep}\rfloor$, the count in the moving unit interval $[n_g,n_g+1)$ satisfies
\begin{equation}
 B_{n_g}^s=\frac1\pi\sqrt{\frac{\beta}{2-\beta}}\log g+O(1)
 \longrightarrow\infty.
 \label{eq:movingbin}
\end{equation}
For $1<\beta<2$, the separatrix instead lies at negative shifted energies for large $g$. Equation~\eqref{eq:betacrowd} applies there as a statement about the full spectrum.

\subsection{Comparison with the lower channel}
To connect the adiabatic potential to the exact spectrum, set
\begin{equation}
 R(q)=\sqrt{\Delta^2+2g^2q^2},\qquad
 \theta(q)=\arctan\frac{\sqrt2gq}{\Delta},
 \label{eq:rotationdefs}
\end{equation}
and rotate the spin by $U=e^{-i\theta\sigma_y/2}$. The scalar band operators are
\begin{equation}
 L_\pm=-\frac12\partial_q^2+\frac{q^2}{2}-\frac12\pm R(q).
 \label{eq:scalarbands}
\end{equation}
The rotation preserves the half-line parity form domains because $U(0)=I$. In the lower channel, positive parity gives the Dirichlet condition and negative parity gives the Neumann condition. Write $N_{-,s}$ for the corresponding strict scalar count.

For $e=e_{\rm sep}+\xi$ with $\xi$ in a compact interval, the upper channel is separated by a gap of order $g^2$. The derivatives of the rotation angle are of orders $g^{-1}$ and $g^{-2}$. 
These derivatives enter the couplings between the two scalar channels. The large gap to the upper channel allows their effect to be bounded by shifts of order $g^{-2}$ in the lower-channel energy threshold.
Completing the block quadratic
form and applying the min–max principle gives
\begin{equation}
 N_{-,s}(e-Cg^{-2})\le\N_s(e)\le N_{-,s}(e+Cg^{-2}).
 \label{eq:betaschurbracket}
\end{equation}
The comparison is uniform on $\mathcal B\times\mathcal K$ and holds also when $e$ is an eigenvalue, with strict counts throughout. Appendix~\ref{app:lower-comparison} gives the form-domain justification and the estimates on the eliminated upper channel.

\subsection{The logarithmic action and quantum count}
The scalar momentum is
\begin{equation}
 p_{\beta,g}(q,\xi)^2
 =2[e_{\rm sep}+\xi-V_-(q)]
 =2\xi+q^2\left(\frac{4g^2}{R(q)+\Delta}-1\right).
 \label{eq:betamomentum}
\end{equation}
Choose a fixed $Q$ large enough and a fixed $\eta>0$ small enough, both uniformly for $\beta\in\mathcal B$. For $Q\le q\le\eta g$,
\begin{equation}
 p_{\beta,g}(q,\xi)^2
 =\kappa_\beta^2q^2+2\xi
 +O_{\mathcal B,\mathcal K}\!\left(\frac{q^2}{g^2}+\frac{q^4}{g^2}\right),
 \label{eq:betainnerexp}
\end{equation}
and hence
\begin{equation}
 \left|\frac1{p_{\beta,g}(q,\xi)}-\frac1{\kappa_\beta q}\right|
 \le C\left(q^{-3}+\frac1{g^2q}+\frac q{g^2}\right).
 \label{eq:betatraveltimeinner}
\end{equation}
The right-hand side is integrable on $[Q,\eta g]$ with a bound independent of $g,\beta,\xi$.

The unique outer turning point satisfies
\begin{equation}
 q_t=2g\sqrt{2-\beta}+O(g^{-1})
 \label{eq:betaturning}
\end{equation}
uniformly on $\mathcal B$. On the remaining interval $[\eta g,q_t]$, the scaling $q=gy$ gives
\begin{equation}
 g^{-2}p_{\beta,g}(gy,\xi)^2
 =P_\beta(y)+O(g^{-2}),
 \qquad
 P_\beta(y)=-2\beta-y^2+2\sqrt{\beta^2+2y^2},
 \label{eq:Pbeta}
\end{equation}
whose outer zero $2\sqrt{2-\beta}$ is simple uniformly on $\mathcal B$. Thus the scaled travel-time integral is uniformly bounded. For the exterior action
\begin{equation}
 I_{\beta,g}(\xi)=\int_Q^{q_t(g,\beta,\xi)}p_{\beta,g}(q,\xi)\,\dd q
 \label{Iint}
\end{equation}
we obtain
\begin{equation}
 I_{\beta,g}'(\xi)=\int_Q^{q_t}\frac{\dd q}{p_{\beta,g}(q,\xi)}
 =\frac1{\kappa_\beta}\log g+O(1),
 \label{eq:betaactionderivative}
\end{equation}
uniformly for $\beta\in\mathcal B$ and $\xi\in\mathcal K$.
The logarithm comes from the increasing range $Q\leq q\leq\eta g$, where the inverse classical momentum is proportional to $1/q$. The outer part of the orbit contributes only a bounded travel time.

A local density estimate alone would not control the number of levels in an individual unit interval. The needed bounded remainder comes from scalar oscillation. The fixed interval $0\leq q\leq Q$ contributes only $O(1)$ zeros and action, so the exterior action \eqref{Iint} determines the growing part of the count. Appendix~\ref{app:lower-oscillation} controls the remaining phase error, including the outer turning-point region. For either boundary condition it gives

\begin{equation}
 N_{-,s}(e_{\rm sep}+\xi)=\frac{I_{\beta,g}(\xi)}{\pi}+O(1),
 \label{eq:betascalarcount}
\end{equation}
uniformly for $\beta\in\mathcal B$ and $\xi\in\mathcal K$.

Take these estimates on a fixed compact enlargement of $\mathcal K$. By \eqref{eq:betaactionderivative}, the threshold displacements in \eqref{eq:betaschurbracket} change the action by only $O(g^{-2}\log g)$. Hence the full quantum count satisfies
\begin{equation}
 \N_s(e_{\rm sep}+\xi)=\frac{I_{\beta,g}(\xi)}{\pi}+O(1).
 \label{eq:betaquantumcount}
\end{equation}
Subtracting at two thresholds gives \eqref{eq:betacrowd}. The uniform remainder permits endpoints that depend on $g$. In particular, the endpoints of $[n_g,n_g+1)$ relative to the separatrix are $n_g-x_{\rm sep}$ and $n_g+1-x_{\rm sep}$. They remain in $[-1,1]$ and have separation one, giving \eqref{eq:movingbin}.
\FloatBarrier

\section{Exact restoration of Braak's counting rules}\label{sec:tail}
The exact tail bound follows from the count at each integer endpoint. For a fixed integer $N\ge0$, define
\begin{equation}
 C_N^s(g,\Delta)=\#\{j:E_j^s<N-g^2\}.
 \label{eq:CN}
\end{equation}

For every $g>0$, real $\Delta$, integer $N\ge0$, and parity $s=\pm$,
\begin{equation}
 N\ge g^2+|\Delta|
 \quad\Longrightarrow\quad
 C_N^s-N\in\{0,1\}.
 \label{eq:binary}
\end{equation}
Braak's counting rules follow by taking differences at consecutive integer endpoints. The argument uses two properties of exceptional levels. Non-Juddian pole eigenvalues are excluded at or above the sufficient boundary, while every Juddian level at pole $N$ has ordered index $N$ in each parity sector. Thus only this level can cross the edge as $\Delta$ varies from zero. We establish these properties using the exceptional-spectrum classification and the matching and determinant formulas of Kimoto, Reyes-Bustos and Wakayama \cite{KRW}.

At a shifted integer pole $x=N$, the exceptional energy is
\begin{equation}
 E=N-g^2,
 \qquad N\in\mathbb Z_{\ge0}.
 \label{eq:poleE}
\end{equation}
The exceptional spectrum separates into the Juddian (polynomial, doubly degenerate) and non-Juddian (non-degenerate) sectors \cite{LiBatchelor2015,LiBatchelor2016,KRW}. Kimoto, Reyes-Bustos and Wakayama characterize the latter by a transcendental constraint function $T_0^{(N)}(g,\Delta)$ \cite{KRW}.

\subsection{Excluding non-Juddian pole eigenvalues}

For $\Delta>0$, we show that the Frobenius solution remains in a region where its two components cannot satisfy the required matching condition at the midpoint.
In the Bargmann representation, use the $\sigma_x$ eigenbasis, remove the factor $e^{-gz}$ and set $y=(z+g)/(2g)$. At \eqref{eq:poleE}, the symmetric Rabi model equations become
\begin{align}
 y\phi_+'&=N\phi_+-\Delta\phi_-,\label{eq:frob1}\\
 (y-1)\phi_-'&=(N-4g^2+4g^2y)\phi_- -\Delta\phi_+ .\label{eq:frob2}
\end{align}
The largest-exponent Frobenius solution is normalized by
\begin{equation}
 \phi_-(y)=y^{N+1}[1+O(y)],
 \qquad
 \phi_+(y)=\frac{N+1}{\Delta}y^N[1+O(y)].
 \label{eq:frobnorm}
\end{equation}
Set
\begin{equation*}
 \rho(y)=\frac{\phi_-(y)}{\phi_+(y)}
 =\frac{\Delta}{N+1}y+O(y^2).
\end{equation*}
Then
\begin{equation}
 y(1-y)\rho'
 =\Delta y+[4g^2y(1-y)-N]\rho+\Delta(1-y)\rho^2.
 \label{eq:riccati}
\end{equation}

For $g,\Delta>0$, the exclusion condition is
\begin{equation}
 N\ge g^2+\Delta.
 \label{eq:highbandcond}
\end{equation}
Under this condition we shall obtain
\begin{equation}
 T_0^{(N)}(g,\Delta)>0.
 \label{eq:Tpos}
\end{equation}
Consequently $E=N-g^2$ cannot be a non-Juddian exceptional eigenvalue.

First suppose $N>g^2+\Delta$ and put
\begin{equation*}
 \alpha=\frac{\Delta}{N-g^2}\in(0,1).
\end{equation*}
Let $F(y,\rho)$ denote the right-hand side of \eqref{eq:riccati}. Then
\begin{equation*}
 F(y,0)=\Delta y>0.
\end{equation*}
At $\rho=\alpha$, using $4y(1-y)\le1$,
\begin{align*}
 F(y,\alpha)
 &\le \Delta y+(g^2-N)\alpha+\Delta(1-y)\alpha^2\\
 &=-\Delta(1-y)(1-\alpha^2)<0.
\end{align*}
Because \eqref{eq:frobnorm} starts inside the strip, $0<\rho<\alpha$ is forward invariant on $0<y\le1/2$. A zero of $\phi_+$ cannot occur there. Boundedness of $\rho$ would force $\phi_-$ to vanish at the same ordinary point, contradicting uniqueness of the nonzero Frobenius solution.

At the boundary $N=g^2+\Delta$, the upper barrier is $\alpha=1$ and
\begin{equation*}
 F(y,1)=-g^2(1-2y)^2<0
 \qquad(0<y<1/2).
\end{equation*}
If $\rho(1/2)=1$, analyticity of \eqref{eq:riccati} gives
\begin{equation*}
 \rho(y)=1-\frac{16g^2}{3}\left(y-\frac12\right)^3
 +O\!\left((y-\tfrac12)^4\right),
\end{equation*}
which is $>1$ immediately to the left of $1/2$, again a contradiction. Hence in all cases
\begin{equation}
 0<\rho(1/2)<1.
 \label{eq:rhomid}
\end{equation}

For the symmetric Rabi model, the KRW midpoint matching factorizes as \cite[Sec.~5.1, Eqs.~(5.11)--(5.13) and Remark~5.1]{KRW}
\begin{equation}
 T_0^{(N)}=(\bar R-R)(\bar R+R).
 \label{eq:Tfact0}
\end{equation}
The unit coefficient of $y^{N+1}$ in \eqref{eq:frobnorm} agrees with their Frobenius normalization in Eqs.~(5.1)--(5.2) and (5.5)--(5.6). Their midpoint definitions give $\bar R^{(N,-)}=\phi_+(1/2)$ and $R^{(N,-)}=\phi_-(1/2)$, while reflection symmetry at zero bias gives $\bar R^{(N,+)}=\bar R^{(N,-)}$ and $R^{(N,+)}=R^{(N,-)}$. Thus their Eqs.~(5.12)--(5.13) yield
\begin{equation}
 T_0^{(N)}
 =\phi_+(1/2)^2\,[1-\rho(1/2)^2]
 \label{eq:Tfact}
\end{equation}
with no additional factor. By \eqref{eq:rhomid} this is positive, giving \eqref{eq:Tpos}. Since the zeros of $T_0^{(N)}$ are precisely the non-Juddian exceptional solutions at \eqref{eq:poleE}~\cite{KRW}, none can occur in this region.

For $N>g^2+\Delta$ the same argument gives the quantitative lower bound
\begin{equation*}
 T_0^{(N)}>
 \left[1-\left(\frac{\Delta}{N-g^2}\right)^2\right]\phi_+(1/2)^2.
\end{equation*}

\subsection{The ordered index of a Juddian level}
For the symmetric model, set
\begin{equation*}
 X=(2g)^2,
 \qquad Y=\Delta^2.
\end{equation*}
For $\Delta>0$ and $N\ge1$, the Juddian locus at pole $N$ is
\begin{equation*}
 P_N^{(N,0)}(X,Y)=0.
 \end{equation*}
The symmetric determinant representation of Kimoto, Reyes-Bustos and Wakayama 
~\cite[Corollary~3.4]{KRW}, specialized to zero bias, is
\begin{equation}
 P_N^{(N,0)}(X,Y)
 =\det\!\left(YI_N+D_NX+S_N\right),
 \qquad D_N=\diag(1,2,\ldots,N),
 \label{eq:constraintdet}
\end{equation}
where the nonzero entries of the real symmetric matrix $S_N$ are \cite[Proposition~3.2 and Corollary~3.4]{KRW}
\begin{equation}
 (S_N)_{ii}=-i[2(N-i)+1],\qquad
 (S_N)_{i,i+1}=(S_N)_{i+1,i}=(N-i)\sqrt{i(i+1)}.
 \label{eq:SNentries}
\end{equation}
The diagonal formula holds for $1\le i\le N$ and the off-diagonal formula for $1\le i<N$.
In particular, $S_N$ is independent of both $X$ and $Y$. For fixed $Y$, the $X$-roots are the eigenvalues of
\begin{equation}
 A_N(Y)=-D_N^{-1/2}(YI_N+S_N)D_N^{-1/2}.
 \label{eq:AN}
\end{equation}
Because $S_N$ is irreducible tridiagonal, these eigenvalues are simple. We label them
\begin{equation*}
 X_1(Y)<X_2(Y)<\cdots<X_N(Y).
\end{equation*}
They are analytic functions of $Y$, and
\begin{equation*}
 A_N'(Y)=-D_N^{-1}.
\end{equation*}
Hellmann--Feynman therefore gives
\begin{equation}
 -1\le X_j'(Y)=-\ip{v_j}{D_N^{-1}v_j}\le-\frac1N<0,
 \label{eq:rootmonotone}
\end{equation}
where $\norm{v_j}=1$. Hence the finite-Jacobi Juddian branches never turn back or intersect.

Positivity at $Y=0$ also follows directly from this matrix. Reversing the basis in $A_N(0)$ gives
$B D_N B^{\mathsf T}$, where $B$ is lower bidiagonal with diagonal entries $1$ and subdiagonal entries $-1$.
Thus $A_N(0)$ is positive definite. The upper bound in \eqref{eq:rootmonotone} ensures that each
$X_j(Y)$ reaches zero at finite $Y$. At the opposite axis,
\begin{equation}
 P_N^{(N,0)}(0,Y)=\prod_{m=1}^N(Y-m^2).
 \label{eq:x0roots}
\end{equation}
The zeros on the $Y$-axis are simple. Since every $X_j(Y)$ decreases strictly, each of the $N$
positive branches issuing from $Y=0$ reaches $X=0$ exactly once. Their ordering gives
$X_j(j^2)=0$, and the endpoints are
\begin{equation}
 g=0,
 \qquad \Delta=m,
 \qquad 1\le m\le N.
 \label{eq:endpoint}
\end{equation}
At \eqref{eq:endpoint}, the decoupled Hamiltonian is $a^\dagger a+m\sigma_z$. The energy $E=N$ is carried by
\begin{equation}
 |N-m,\uparrow\rangle,
 \qquad
 |N+m,\downarrow\rangle,
 \label{eq:endstates}
\end{equation}
which have opposite parity. Put $A=N-m$ and $B=N+m$. These integers have the same parity.
If $s=(-1)^A$ is the parity of the first state in \eqref{eq:endstates}, the strict counts below $N$ in parity $s$ and parity $-s$ are, respectively,
\begin{equation}
 \left\lfloor\frac A2\right\rfloor
 +\left\lceil\frac B2\right\rceil
 =\left\lceil\frac A2\right\rceil
 +\left\lfloor\frac B2\right\rfloor=N.
 \label{eq:endcount1}
\end{equation}
Thus the pole state has zero-based ordered index $N$ in each parity chain at the endpoint.
Along a connected Juddian branch, the Hamiltonian and its parity-resolved eigenprojections vary
continuously. Since each fixed-parity Jacobi operator has simple spectrum, the ordered index cannot
change without a same-parity degeneracy. It therefore remains equal to $N$ everywhere on the branch.

At every Juddian point on the pole $E=N-g^2$,
\begin{equation}
 E_N^+(g,\Delta)=E_N^-(g,\Delta)=N-g^2.
 \label{eq:index}
\end{equation}
With strict counting this reads
\begin{equation*}
 \N_+(N-g^2)=\N_-(N-g^2)=N.
\end{equation*}

This is the ordered-level form of the statement that $P_N^{(N,0)}(X,Y)=0$ gives the $N$th degenerate pair \cite{LiFerriBatchelor2021}.

\subsection{Counts at the integer endpoints}
We now obtain \eqref{eq:binary} by varying the atomic splitting. The resulting counting inequalities are, for every integer $n\ge g^2+|\Delta|$,
\begin{equation}
 B_n^s\in\{0,1,2\},
 \qquad
 B_n^s+B_{n+1}^s\in\{1,2,3\}.
 \label{eq:tailbraak}
\end{equation}
Thus every violation of these counting inequalities is confined to
\begin{equation}
 n<g^2+|\Delta|.
 \label{eq:violationregion}
\end{equation}
This threshold is sufficient and need not be optimal for a given parameter pair.

It is enough to take $\Delta\ge0$, since changing the sign of $\Delta$ unitarily exchanges the parity sectors. Fix $g,N$ and vary $\Delta'$ continuously from $0$ to $\Delta$, staying in
\begin{equation*}
 0\le\Delta'\le\Delta\le N-g^2.
\end{equation*}
At $\Delta'=0$ the two parity spectra are displaced oscillator ladders, so
\begin{equation*}
 C_N^s(g,0)=N.
\end{equation*} 
At $\Delta'=0$, the difference between the $j$th eigenvalue and the pole is $j-N$. The ordered eigenvalues depend continuously on $\Delta'$. A branch with $j\ne N$ could reach the pole only at a positive value of $\Delta'$. Such a meeting cannot be non-Juddian by \eqref{eq:Tpos}, while \eqref{eq:index} assigns index $N$ to every Juddian eigenvalue at this pole. Hence the first $N$ branches remain strictly below the pole and all branches with $j>N$ remain strictly above it. Only $E_N^s$ can lie on either side.
Therefore
\begin{equation*}
 C_N^s=N+\mathbf 1_{\{E_N^s<N-g^2\}},
\end{equation*}
giving \eqref{eq:binary}. At a Juddian crossing the indicator is zero in both parities, as required by strict counting. At $\Delta=0$, the exact oscillator ladder gives $B_n^s=1$ directly.

Put $d_N^s=C_N^s-N\in\{0,1\}$. Then
\begin{equation*}
 B_n^s=C_{n+1}^s-C_n^s
 =1+d_{n+1}^s-d_n^s\in\{0,1,2\},
\end{equation*}
and
\begin{equation*}
 B_n^s+B_{n+1}^s
 =C_{n+2}^s-C_n^s
 =2+d_{n+2}^s-d_n^s\in\{1,2,3\}.
\end{equation*}

The bounded perturbation estimate (17) also gives a
coupling-independent validity range. For $|\Delta|<1$,
it places $x_j^s$ strictly between $j-1$ and $j+1$.
Consequently,
\[
C_N^s=N+\mathbf{1}_{\{x_N^s<N\}}
\]
for every $N\geq0$. Taking differences establishes
the counting rules in every nonnegative unit interval.
The endpoint $|\Delta|=1$ is also included for $g>0$.
Indeed, the Jacobi recurrence implies that every eigenvector has nonzero components of both oscillator parities.
The Hellmann-Feynman formula therefore gives
$|\partial_\Delta E_j^s|<1$, making the displacement
bound (17) strict for $\Delta\neq0$.
Thus the counting rules hold globally for
$|\Delta|\leq1$.

\section{Macroscopic spectral density}\label{sec:density}
\begin{figure}[htbp]
\centering
\includegraphics[width=0.96\linewidth]{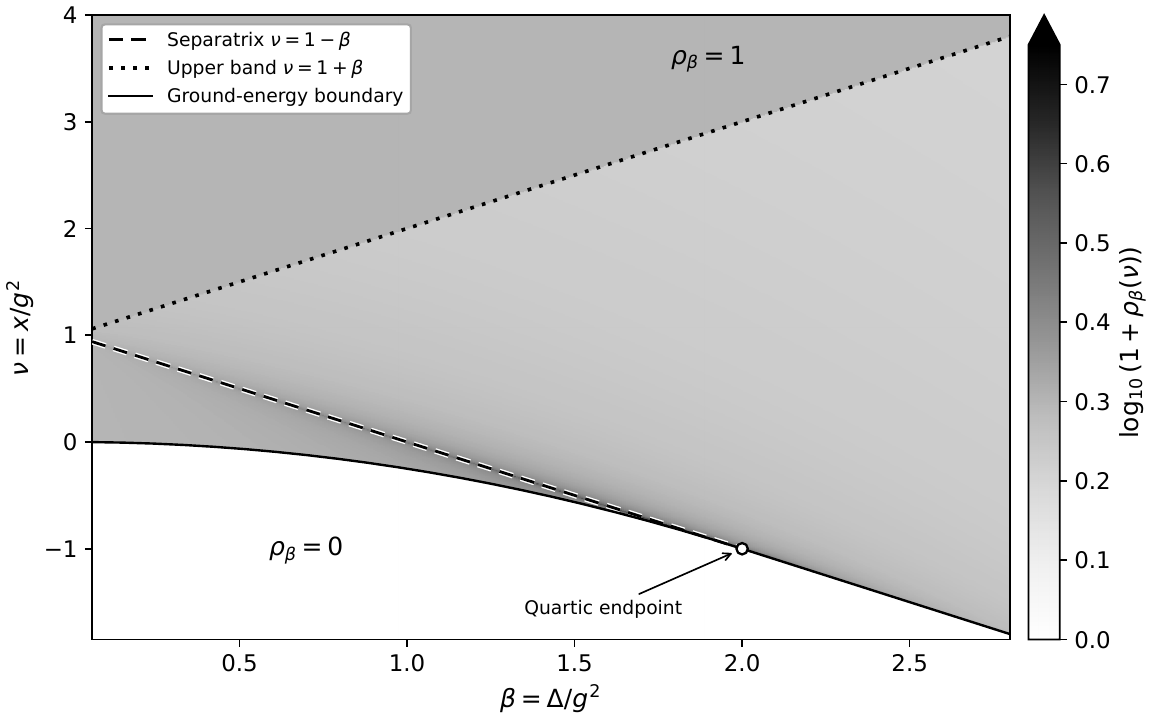}
\caption{Macroscopic same-parity spectral density in the joint scaling \eqref{eq:macroscale}. The dashed line is the lower separatrix $\nu=1-\beta$ for $0<\beta<2$. The dotted line is the upper-band onset $\nu=1+\beta$. Above it the density is exactly $1$. The grey scale shows $\log_{10}(1+\rho_\beta)$ and is capped to display the separatrix ridge. The solid lower curve is the classical ground-energy boundary.}
\label{fig:density}
\end{figure}

To place the separatrix count in the full spectrum, let $\beta>0$ denote the limiting ratio $\Delta/g^2$ and take
\begin{equation}
 \Delta=\beta g^2+o(g^2),
 \qquad
 x=\nu g^2+O(1),
 \qquad g\to\infty.
 \label{eq:macroscale}
\end{equation}
With $q=gy$, the two adiabatic surfaces are
\begin{equation}
 w_\pm(y)=\frac{y^2}{2}\pm\sqrt{\beta^2+2y^2}.
 \label{eq:bands}
\end{equation}
The lower edge of the classical spectrum is
\begin{equation}
 \nu_{\min}=\begin{cases}-\beta^2/4,&0<\beta\le2,\\1-\beta,&\beta\ge2.\end{cases}
 \label{eq:macroground}
\end{equation}
Figure~\ref{fig:density} shows the resulting spectral regions. The lower separatrix is $\nu=1-\beta$ for $0<\beta<2$, while the upper band opens at $\nu=1+\beta$. The two lines and the quartic endpoint organize the density calculation below.

For $\beta>0$, define the half-line actions
\begin{equation}
 I_\pm(\nu)=\int_0^\infty\sqrt{2\{\nu-1-w_\pm(y)\}_+}\,\dd y.
 \label{eq:macroactions}
\end{equation}
Here $z_+=\max(z,0)$. The leading semiclassical counting law is
\begin{equation}
 g^{-2}\N_s((\nu-1)g^2)\longrightarrow
 \frac{I_-(\nu)+I_+(\nu)}{\pi}.
 \label{eq:macrocount}
\end{equation}
To justify \eqref{eq:macrocount}, rescale with $q=gy$ and set $h=g^{-2}$ and $\beta_g=\Delta/g^2\to\beta>0$. The scaled matrix operator is
\begin{equation}
 g^{-2}H=-\frac{h^2}{2}\partial_y^2+\frac{y^2}{2}
 +\sqrt2y\,\sigma_x+\beta_g\sigma_z-\frac h2.
 \label{eq:macroscaledoperator}
\end{equation}
The rotation angle $\arctan(\sqrt2y/\beta_g)$ has uniformly bounded first and second derivatives. For a rotated spinor $f$, Cauchy--Schwarz bounds its derivative coupling, in quadratic-form sense, by
$Ch\{h^2\|f'\|^2+\|f\|^2\}$. The rotated operator is therefore bracketed by direct sums of scalar band operators with kinetic coefficients $1+O(h)$ and energy shifts $O(h)$. Moreover,
$|w_\pm(y;\beta_g)-w_\pm(y;\beta)|\le|\beta_g-\beta|$ uniformly in $y$. The confining scalar Weyl law and the min--max principle give \eqref{eq:macrocount}, since each band contributes its half-line action divided by $\pi h$. The action is continuous even at a critical energy, so the vanishing threshold shifts do not affect this leading limit. Parity changes the boundary condition, which also leaves the leading term unchanged. The full spectrum has twice the leading count in \eqref{eq:macrocount}. Differentiation at regular energies defines the macroscopic density
\begin{equation}
 \rho_\beta(\nu)=\rho_{\beta,-}(\nu)+\rho_{\beta,+}(\nu),
 \label{eq:rho}
\end{equation}
with
\begin{equation}
 \rho_{\beta,\mp}(\nu)=\frac1\pi
 \int_{\mathcal A_\mp(\beta,\nu)}
 \frac{\dd y}{\sqrt{2\{\nu-1-y^2/2\pm\sqrt{\beta^2+2y^2}\}}}.
 \label{eq:rhopm}
\end{equation}
Here $\mathcal A_\pm(\beta,\nu)=\{y\ge0:w_\pm(y)<\nu-1\}$. This is a density of the limiting counting measure. The counts in individual unit intervals of shifted energy remain integers and are not asserted to converge pointwise to $\rho_\beta(\nu)$.

For $0<\beta<2$ the lower surface has its separatrix at
\begin{equation}
 \nu_{\rm sep}=1-\beta,
\end{equation}
while the upper band opens at
\begin{equation}
 \nu_+=1+\beta.
\end{equation}
Near the lower separatrix,
\begin{equation}
 \rho_\beta(\nu)
 =\frac{1}{2\pi\kappa_\beta}
 \log\frac1{|\nu-(1-\beta)|}+O(1).
 \label{eq:rhosing}
\end{equation}
The coefficient agrees with the counting law for a fixed interval because a unit interval in shifted energy has macroscopic width $g^{-2}$, giving $\log(g^2)=2\log g$. The uniform remainder for the quantum count was established independently in Section~\ref{sec:crowding}.

Above the upper-band onset, for every $\beta>0$,
\begin{equation}
 \rho_\beta(\nu)=1,
 \qquad \nu>1+\beta.
 \label{eq:rhoone}
\end{equation}
At the onset itself the density in this statement is understood as its right-hand limit.

Let $\varepsilon=\nu-1\ge\beta$ and consider the two full-line classical symbols
\begin{equation}
 h_\pm(y,p)=\frac{p^2+y^2}{2}\pm\sqrt{\beta^2+2y^2}.
\end{equation}
Put $y=R\cos\theta$, $p=R\sin\theta$, and $u=R^2$. Along a fixed ray, the two band crossings satisfy
\begin{equation}
 \left(\varepsilon-\frac{u}{2}\right)^2=\beta^2+2u\cos^2\theta.
\end{equation}
If $u_1,u_2$ are the roots, then
\begin{equation}
 u_1+u_2=4(\varepsilon+2\cos^2\theta).
\end{equation}
Because both bands are occupied at the origin when $\varepsilon\ge\beta$, the multiplicity-weighted radial area is $(u_1+u_2)/2$. Integration over $\theta$ gives the full-line phase volume
\begin{equation}
 \Omega(\varepsilon)=4\pi(\varepsilon+1)=4\pi\nu.
\end{equation}
Reflection in coordinate and momentum gives four copies of each half-line action in the full phase volume. Hence
\begin{equation}
 I_-(\nu)+I_+(\nu)=\pi\nu.
\end{equation}
Differentiation for $\varepsilon>\beta$ gives \eqref{eq:rhoone}.

The numerical evaluation of Figure~\ref{fig:density} is described in Appendix~\ref{app:numerics-density}.

The sufficient quantum tail boundary \eqref{eq:tailbraak} becomes, under $n=\nu g^2$ and $\Delta=\beta g^2$,
\begin{equation}
 \nu\ge1+\beta.
\end{equation}
At finite $g$, the upper adiabatic minimum has shifted energy $g^2+\Delta-1/2$, whereas the sufficient tail condition is $n\ge g^2+\Delta$. Their macroscopic limits coincide. The exact quantum restoration follows from Section~\ref{sec:tail}.

\FloatBarrier

\section{The quartic crossover}\label{sec:quartic}
The hyperbolic curvature \eqref{eq:kappa} vanishes as $\beta\to2^-$. The quartic critical potential and finite-frequency scaling are established features of the Rabi transition \cite{Hwang2015,Puebla}. Here we derive a counting law with a uniform $O(1)$ remainder across bounded crossover windows. 
Near the endpoint, the lower potential has the expansion
\[
V_-(q)-e_{\mathrm{sep}}
=
\left(\frac{1}{2}-\frac{g^2}{\Delta}\right)q^2
+\frac{g^4}{2\Delta^3}q^4
+O\left(\frac{g^6q^6}{\Delta^5}\right).
\]
The quadratic curvature vanishes at $\Delta=2g^2$.
When $2g^2-\Delta=O(g)$, the quadratic and quartic
terms are both of order one for $q=O(\sqrt{g})$.
This motivates the scaling
\begin{equation}
 \lambda=g\left(2-\frac{\Delta}{g^2}\right),\qquad q=\sqrt g\,y,\qquad h=g^{-1/2}.
 \label{eq:quarticscale}
\end{equation}
The limiting lower potential, measured from its value at the origin, is
\begin{equation}
 U_\lambda(y)=\frac{y^4}{16}-\frac{\lambda y^2}{4}.
 \label{eq:quarticU}
\end{equation}
Positive $\lambda$ gives a barrier at the half-line boundary and a well at $y=\sqrt{2\lambda}$. At $\lambda=0$ the minimum is quartic, while negative $\lambda$ gives a single quadratic minimum at the boundary.

Write
\begin{equation}
 I_\lambda(\xi)=\int_0^\infty\sqrt{2(\xi-U_\lambda(y))_+}\,\dd y,
 \qquad
 \mathcal F_\lambda(a,b)=\frac{I_\lambda(b)-I_\lambda(a)}{\pi}.
 \label{eq:Fquartic}
\end{equation}
For $\Delta=2g^2-\lambda g$ and $e_{\rm sep}=-\Delta-1/2$, let
$\W_s(a,b;g)=\#\{j:a\le E_j^s-e_{\rm sep}<b\}$. The universal counting law is
\begin{equation}
 \W_s(a,b;g)=\sqrt g\,\mathcal F_\lambda(a,b)+O(1).
 \label{eq:quarticcount}
\end{equation}
For every fixed $L$ and compact energy interval $\mathcal K$, the remainder is uniform for $|\lambda|\le L$, either parity, and $a<b$ in $\mathcal K$. The reference $e_{\rm sep}$ continues to be used when $\lambda\le0$, although the origin is then a minimum rather than a barrier.

The shifted reference energy in this scaling is
\begin{equation}
 x_{\rm sep}=-g^2+\lambda g-\frac12.
 \label{eq:quarticxsep}
\end{equation}
For bounded $\lambda$ and bounded $a,b$, the entire window $[x_{\rm sep}+a,x_{\rm sep}+b)$ therefore lies at negative shifted energy for sufficiently large $g$. This crossover describes the endpoint of the crowding mechanism in the full spectrum. Violations of the counting rules in the intervals $[n,n+1)$ with $n\ge0$ are supplied by the logarithmic regime of Section~\ref{sec:crowding}.

\subsection{The universal function}
For $\xi>0$, differentiation of the quartic action gives
\begin{equation}
 \rho_\lambda(\xi)=\frac{I_\lambda'(\xi)}{\pi}
 =\frac1\pi\sqrt{\frac{2}{\chi}}
 K\!\left(\sqrt{\frac{\chi+\lambda}{2\chi}}\right),\qquad
 \chi=\sqrt{\lambda^2+4\xi},
 \label{eq:ellipticrho}
\end{equation}
where $K$ is the complete elliptic integral with the modulus as its argument. 
At the endpoint, writing $\mathrm B$ for the beta function, we obtain
\begin{equation}
 \mathcal F_0(a,b)=\frac{\mathrm B(1/4,3/2)}{\pi\sqrt2}
 \left[(b_+)^{3/4}-(a_+)^{3/4}\right].
 \label{eq:purequartic}
\end{equation}
Equation \eqref{eq:purequartic} makes the $\sqrt g$ growth explicit at the quartic endpoint.

\begin{figure}[t]
\centering
\includegraphics[width=0.88\linewidth]{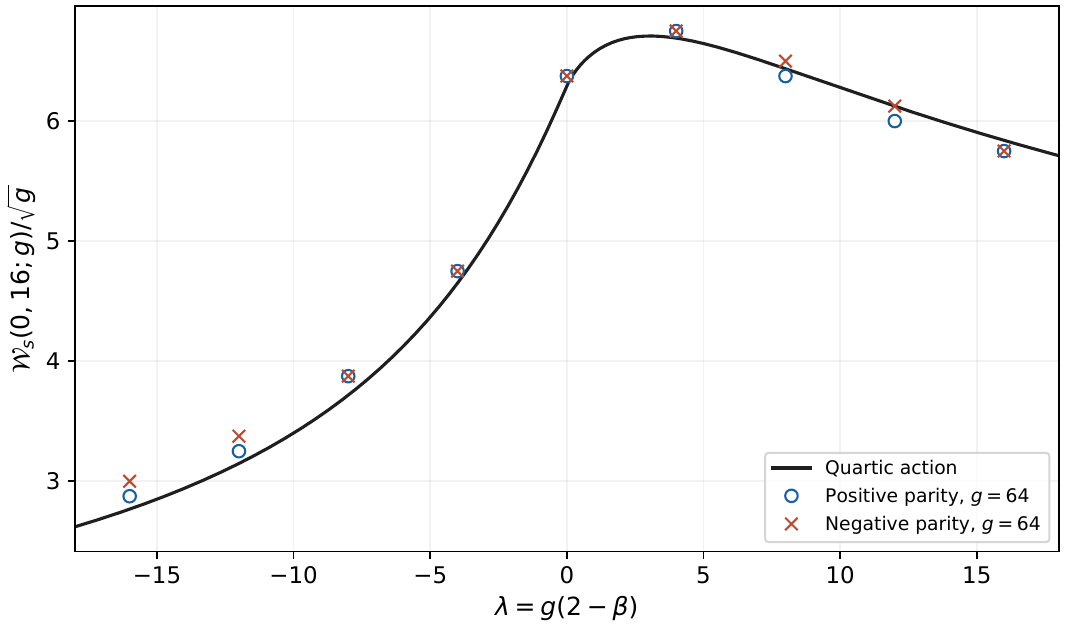}
\caption{Universal quartic crossover in the barrier-relative window $[0,16)$. The curve is $\mathcal F_\lambda(0,16)$ from \eqref{eq:Fquartic}. Symbols are full parity-Jacobi counts divided by $\sqrt g$ at $g=64$, with $\Delta=2g^2-\lambda g$. No fitted parameters are used.}
\label{fig:quartic}
\end{figure}

Figure~\ref{fig:quartic} compares the universal action with the full parity-Jacobi counts. As an additional example,
\begin{equation}
 \mathcal F_2(-2,2)=2.2377123\ldots.
 \label{eq:quarticexample}
\end{equation}
At $g=20,40,80$, the positive-parity counts divided by $\sqrt g$ are $2.2361$, $2.2136$ and $2.2361$, respectively. These finite-section counts are checked at two truncation dimensions, as described in Appendix~\ref{app:numerics-quartic}.

The universal function also has the expected outer limits. For fixed $\xi>0$, \eqref{eq:ellipticrho} gives
\begin{equation}
 \rho_\lambda(\xi)\sim
 \frac1\pi\sqrt{\frac2\lambda}\log\frac{4\lambda}{\sqrt \xi}
 \quad(\lambda\to+\infty),\qquad
 \rho_\lambda(\xi)\sim\frac1{\sqrt{2|\lambda|}}
 \quad(\lambda\to-\infty).
 \label{eq:quarticouterlimits}
\end{equation}
These are the hyperbolic and half-line harmonic limits of the universal function. The quantum estimate \eqref{eq:quarticcount} is uniform only for bounded $\lambda$. No uniform double-scaling overlap with the fixed-$\beta$ law is claimed here.

\subsection{Connection with the quantum count}
The reduction and uniform estimates are given in Appendix~\ref{app:quartic}. The upper adiabatic channel remains separated from the counting window by an energy of order $g^2$. Eliminating it brackets the exact parity count between lower-channel scalar counts with threshold shifts of order $g^{-2}$. After $q=\sqrt g\,y$, the scalar potential differs from $U_\lambda$ by $O(h^2)$ on a fixed interval containing all classically allowed motion. Dirichlet--Neumann bracketing then reduces the count to the half-line operator $-h^2\partial_y^2/2+U_\lambda(y)$, with the boundary condition inherited from parity.

The remaining issue is uniformity as the saddle and minimum merge. An exact dilation separates the quartic region from a compact family with only nondegenerate critical points. A Sturm oscillation estimate gives the action divided by $\pi h$ with an $O(1)$ remainder throughout this family. The action is uniformly H\"older continuous with exponent $3/4$ in energy, so the $O(h^2)$ threshold shifts change the action count by only $O(h^{1/2})$. 
This yields the cumulative estimate \eqref{eq:quarticcumulative}. Subtracting at $a$ and $b$ gives the universal counting law~\eqref{eq:quarticcount}.


\section{Concluding remarks}\label{sec:discussion}

We have shown that the quantum Rabi spectrum admits arbitrarily strong local crowding while retaining exact counting constraints. Along $\Delta=g^2-1/2$, every fixed interval $[n,n+1)$ in shifted energy, with $n\ge0$, contains $(\log g)/\pi+O_n(1)$ levels of either parity. At arbitrary parameters the cumulative parity counts differ by at most one, and Braak's counting rules hold for $n\ge g^2+|\Delta|$. 
The threshold is sufficient and need not be optimal for a given parameter pair.
The counting rules also hold in every nonnegative unit interval for all $g>0$ when $|\Delta| \le 1$.
The crowding mechanism is supplied by the lower adiabatic separatrix. Its logarithmically increasing travel time becomes a local counting law for the full operator because the reduction and oscillation estimates have bounded counting errors. The macroscopic density returns to the oscillator value above the upper-band onset. At the quartic endpoint, the corresponding local counts grow as $\sqrt g$ in bounded windows at negative shifted energy.

The joint limit differs from increasing the coupling at fixed atomic splitting. Along the curve \eqref{eq:betascaling}, the atomic frequency is $\Omega=2\Delta\sim2\beta g^2$, so $g/\omega\to\infty$ while $g/\Omega\to0$. The atomic term remains comparable to the displacement energy. Coupling-regime classifications also depend on the accessible energy~\cite{Rossatto2017}. The resonance boundaries and perturbative deep strong coupling description in that classification therefore do not by themselves justify a displaced-oscillator approximation along the present path. Here the fast atomic degree of freedom instead determines the adiabatic surfaces.

At fixed $g,\Delta$, the large-index expansion of Boutet de Monvel and Zielinski is~\cite{BdMZ}
\begin{equation}
 E_n^\pm=n-g^2\pm\frac{\Delta}{\sqrt{2\pi g}}
 \frac{(-1)^n\cos(4g\sqrt n-\pi/4)}{n^{1/4}}
 +O(n^{-1/2+\epsilon}),
 \label{eq:BdMZ}
\end{equation}
for every $\epsilon>0$. Rudnick combined this expansion with estimates from uniform distribution theory to establish the alternating two-level and zero-level parity pattern for all but at most $O(N^{3/4+o(1)})$ integers $n\le N$, at fixed $g,\Delta>0$~\cite[Theorem~1.2]{Rudnick}. Thus the counting rules hold on a set of natural density one. The uncontrolled set may still be infinite. Near zeros of the leading oscillatory factor, the remainder can prevent the asymptotic argument from fixing the signs at integer endpoints. The exceptional-spectrum argument in Section~\ref{sec:tail} establishes the counting inequalities for every $n\ge g^2+|\Delta|$. This complements Rudnick's more detailed description of the typical parity pattern by excluding all violations above an explicit sufficient threshold. Finite-window validity at sufficiently small gap was established by Malagutti and Parmeggiani \cite{MPFerrara,MP2026}.

The analytic level-spacing and consecutive-spacing-ratio results of Ref.~\cite{LiLiSierraBatchelor} use the fixed-parameter high-energy expansion \eqref{eq:BdMZ} in windows containing many spectral oscillations. Their derivation is independent of Braak’s counting inequalities. The splitting parameter in that work equals $2\Delta$ in our notation.

The local counts studied here describe a different limit. Along $\Delta=g^2-1/2$, the fixed interval $[n,n+1)$ in shifted energy corresponds to the physical window $[n-g^2,n+1-g^2)$. The ordered indices of the levels it contains grow, while the sufficient tail boundary recedes as $2g^2-1/2$. These observations reconcile unbounded level counts in fixed intervals with tail restoration at each fixed parameter pair. The counting inequalities alone do not determine spacing distributions~\cite{BraakNguyen}, and their sufficient threshold does not specify when Eq.~\eqref{eq:BdMZ} becomes quantitatively accurate.

Lanuza's proposed eigenvalue inequalities are~\cite{Lanuza2026}
\begin{equation}
 |E_n^+-E_n^-|\le1,\qquad
 E_{n+1}^s-E_n^s\le2,\qquad
 E_n^s\le n+\frac12-g^2.
 \label{eq:lanuzabounds}
\end{equation}
Unbounded interval counts are compatible with these bounds. The parity-count relation controls cross-index ordering and does not establish an energy-difference bound at equal index. Whether the inequalities \eqref{eq:lanuzabounds} hold at arbitrary parameters remains open.

The present results establish unbounded level counts in the intervals $[n,n+1)$ with $n\ge0$, but do not determine the fastest possible growth. Whether another parameter path gives faster growth, or logarithmic growth is optimal, remains open. The quartic law concerns negative shifted energies and does not settle this question. The local spacing statistics in the separatrix and quartic regimes also remain to be determined.

\section*{Acknowledgements}

Murray Batchelor thanks Hai-Qing Lin for hospitality at the Institute for Advanced Study in Physics at Zhejiang University, Hangzhou where this work was completed and presented at the International Workshop on Strongly Coupled Light-Matter Interactions, October 26-28, 2026. 
The work of Zimin Li was supported by the National Natural Science Foundation of China (Grant No. 12205383) and the Hunan Provincial Natural Science Foundation (Grant No. 2024JJ6483).

During the preparation of this manuscript, the authors used ChatGPT Pro to assist with literature searches, mathematical analysis and numerical checks, and the organization and presentation of the manuscript. All mathematical content, references, and figures were independently reviewed and verified by the authors, who take full responsibility for the final content of the article.

\section*{Data availability}
The data that support the findings of this study are available from the authors upon reasonable request.


\section*{Appendices}

\appendix
\numberwithin{equation}{section}
\section{Lower-channel comparison and uniform counting estimates}\label{app:lower}
In this Appendix we give the operator comparison and bounded oscillation estimate used in Section~\ref{sec:crowding}. Throughout, $c$ is fixed, $\beta$ lies in a compact subset $\mathcal B$ of $(0,2)$ and the barrier-relative energy lies in a fixed compact interval $\mathcal K$.

\subsection{Comparison with the lower scalar channel}\label{app:lower-comparison}
Use the rotation angle and scalar operators defined in \eqref{eq:rotationdefs} and \eqref{eq:scalarbands}.
For $g$ large enough, uniformly for $\beta\in\mathcal B$,
\begin{equation}
 \|\theta'\|_\infty\le Cg^{-1},
 \qquad
 \|\theta''\|_\infty\le Cg^{-2}.
 \label{eq:betaderivative}
\end{equation}
After the real spin rotation $U=e^{-i\theta\sigma_y/2}$, the Hamiltonian has the block form
\begin{equation}
 U^\dagger HU=
 \begin{pmatrix}L_++d&K\\K^\dagger&L_-+d\end{pmatrix},
 \qquad
 K=A\partial_q+\frac{A'}2,
 \quad A=\frac{\theta'}2,
 \quad d=\frac{\theta'^2}{8},
 \label{eq:betablock}
\end{equation}
with $L_\pm$ as in \eqref{eq:scalarbands}.
Let
\begin{equation}
 \mathcal Q_N=\{f\in H^1(\R_+):qf\in L^2(\R_+)\},\qquad
 \mathcal Q_D=\{f\in\mathcal Q_N:f(0)=0\}.
 \label{eq:parityformdomains}
\end{equation}
The parity form domains are $\mathcal Q_N\oplus\mathcal Q_D$ for $s=+$ and
$\mathcal Q_D\oplus\mathcal Q_N$ for $s=-$. Since $U(0)=I$ and its derivatives are bounded,
the rotation preserves these domains. For a rotated spinor $(u,f)^{\mathsf T}$, the operator boundary conditions remain
\begin{equation}
 \begin{array}{c|cc}
  s&\text{upper channel }u&\text{lower channel }f\\ \hline
  +&u'(0)=0&f(0)=0\\
  -&u(0)=0&f'(0)=0
 \end{array}
 \label{eq:rotatedbc}
\end{equation}
Indeed, the additional term from $U'(0)$ in either Neumann condition multiplies the component
whose boundary value is zero. Integration by parts gives $K^\dagger=-A\partial_q-A'/2$
with these complementary conditions, since the boundary term $A\overline u f$ vanishes.
Thus $L_{-,+}$ has the Dirichlet condition and $L_{-,-}$ the Neumann condition.

Write $\xi=e-e_{\rm sep}=x-x_{\rm sep}$ for barrier-relative energy. Let
$e=e_{\rm sep}+\xi=-\Delta-1/2+\xi$, with $\xi$ in a fixed compact set. The upper block satisfies
\begin{equation}
 T_e:=L_++d-e\ge c_0g^2I
 \label{eq:betauppergap}
\end{equation}
uniformly on $\mathcal B$. For $f$ in the lower form domain, $Kf\in L^2$ and $T_e^{-1}Kf$ belongs to the upper operator domain.
Completing the square therefore gives, on the parity form domain,
\begin{align}
 \langle(u,f),(U^\dagger HU-e)(u,f)\rangle
 &=\|T_e^{1/2}(u+T_e^{-1}Kf)\|^2+\langle f,S_s(e)f\rangle,\nonumber\\
 S_s(e)&=L_{-,s}+d-e-K^\dagger T_e^{-1}K.
 \label{eq:schurcompletion}
\end{align}
The triangular change of variables is invertible on this form domain. This justifies the
negative-index identity below at every threshold, including eigenvalues.
Equations \eqref{eq:betaderivative}-\eqref{eq:betauppergap} imply
\begin{equation}
 0\le\langle f,K^\dagger T_e^{-1}Kf\rangle
 \le Cg^{-4}\|f'\|^2+Cg^{-6}\|f\|^2.
 \label{eq:betaschurerror}
\end{equation}
For $\beta\in\mathcal B$ the lower potential is bounded below by $-Cg^2$. In fact, because $0<\Delta<2g^2$ for large $g$,
\begin{equation}
 \min V_-=-g^2-\frac12-\frac{\Delta^2}{4g^2}.
\end{equation}
Let $A_s=L_{-,s}+d$ and $D_e=K^\dagger T_e^{-1}K$. Choose $C_1$ so that, as quadratic forms,
\begin{equation}
 A_s+C_1g^2\ge\tfrac12(-\partial_q^2)+1.
\end{equation}
The estimate \eqref{eq:betaschurerror} then gives
\begin{equation}
 0\le D_e\le\varepsilon_g(A_s+C_1g^2),\qquad
 \varepsilon_g=Cg^{-4}.
 \label{eq:schurrelative}
\end{equation}
Since $T_e$ is positive, completing the block form preserves its negative index and gives
\begin{equation}
 \N_s(e)=n_-(S_s(e)),\qquad
 (1-\varepsilon_g)A_s-e-\varepsilon_g C_1g^2
 \le S_s(e)\le A_s-e.
 \label{eq:schurinertia}
\end{equation}
Here $n_-$ counts strictly negative eigenvalues. The effective threshold in the left inequality is
\begin{equation}
 \frac{e+\varepsilon_g C_1g^2}{1-\varepsilon_g}
 =e+O(g^{-2}),
 \label{eq:schurthreshold}
\end{equation}
because $e=O(g^2)$. Finally $0\le d\le Cg^{-2}$, so the min--max principle gives the scalar-count bracket \eqref{eq:betaschurbracket} used in the main text.

\subsection{Uniform control of the scalar count}\label{app:lower-oscillation}
The action estimate must be accompanied by a bounded oscillation remainder. With $P=p_{\beta,g}^2$,
\begin{equation}
 P'=2q\left(\frac{2g^2}{R}-1\right),
 \qquad
 P''=-2+\frac{4g^2\Delta^2}{R^3}.
 \label{eq:betapderivatives}
\end{equation}
On $[Q,\eta g]$ these give $p\asymp q$, $|p'|\le C$, $|p''|\le C/q$.
On the outer allowed region, write $\ell=q_t-q$. Uniform simplicity of the scaled outer root gives
\begin{equation}
 p\asymp\sqrt{g\ell},\qquad
 |p'|\le C\sqrt g\,\ell^{-1/2},\qquad
 |p''|\le C\sqrt g\,\ell^{-3/2}.
 \label{eq:betaouterest}
\end{equation}
After the Liouville change of variable $z=\int p\,\dd q$ and $f=p^{-1/2}u$, the phase error is bounded by
\begin{equation}
 \int\left(\frac{3(p')^2}{4p^3}+\frac{|p''|}{2p^2}\right)\dd q.
 \label{eq:betapruefererror}
\end{equation}
The inner part of \eqref{eq:betapruefererror} is $O(1)$, while \eqref{eq:betaouterest} gives
$O(1)$ down to $\ell=g^{-1/3}$. In the remaining layer $0\le\ell\le g^{-1/3}$ one has
$0\le p^2\le Cg^{2/3}$. Sturm comparison with the constant-coefficient equation of wavenumber
$Cg^{1/3}$ on an interval of length $g^{-1/3}$ therefore gives only $O(1)$ further zeros. Beyond
$q_t$ the scalar equation is disconjugate, so at most one additional zero is possible.

The fixed interval $[0,Q]$ contributes only $O(1)$ zeros and action. Relative to $e_{\rm sep}$ its potential and
the threshold $\xi$ stay in a bounded set, uniformly on $\mathcal B\times\mathcal K$. Sturm oscillation therefore gives \eqref{eq:betascalarcount} for either scalar boundary condition, with a remainder uniform in $\beta\in\mathcal B$ and $\xi\in\mathcal K$.

\section{Uniform estimates for the quartic crossover}\label{app:quartic}
In this Appendix we establish the uniform remainder in \eqref{eq:quarticcount}. Throughout this appendix, $|\lambda|\le L$ and the barrier-relative energy lies in a fixed compact interval $\mathcal K$. All constants may depend on these fixed bounds but are independent of $g$, the parity and the energy within the stated sets.

\subsection{Reduction to the quartic operator}
For bounded $\lambda$ and sufficiently large $g$, one has $g^2\le\Delta=2g^2-\lambda g\le3g^2$. The derivative bounds \eqref{eq:betaderivative} therefore remain uniform. For $e=e_{\rm sep}+\xi$, $\xi\in\mathcal K$, the upper block satisfies $T_e\ge2\Delta-\sup\mathcal K\ge c_0g^2$. Also,
\begin{equation}
 V_-(q)\ge\frac{q^2}{2}-\sqrt2gq-\Delta-\frac12
 \ge-g^2-\Delta-\frac12.
 \label{eq:quarticlowerbound}
\end{equation}
Thus the lower potential is bounded below by $-Cg^2$ on both sides of the quartic endpoint. These are the bounds used in \eqref{eq:schurrelative}--\eqref{eq:schurthreshold}, which consequently give
\begin{equation}
 N_{-,s}(e-Cg^{-2})\le\N_s(e)\le N_{-,s}(e+Cg^{-2}).
 \label{eq:quarticschur}
\end{equation}
Rescaling $q=\sqrt g\,y$ gives the scalar potential
\begin{equation}
 W_{g,\lambda}(y)=\frac{g y^2}{2}+\Delta-\sqrt{\Delta^2+2g^3y^2}.
 \label{eq:Wrquartic}
\end{equation}
For each fixed $Y$, expansion of the square root is uniform on $|\lambda|\le L$, $0\le y\le Y$ and yields
\begin{equation}
 W_{g,\lambda}(y)=U_\lambda(y)+h^2R_{g,\lambda}(y),\qquad
 |R_{g,\lambda}(y)|\le C_{L,Y}.
 \label{eq:quarticpotentialexp}
\end{equation}
There is no remote low-energy well. Indeed,
\begin{equation}
 W_{g,\lambda}'(y)=gy\left(1-\frac{2g^2}{\sqrt{\Delta^2+2g^3y^2}}\right).
 \label{eq:quarticWprime}
\end{equation}
For $\lambda\le0$ this is positive on $y>0$. For $\lambda>0$ its unique positive minimum is at
\begin{equation}
 y_{\min}^2=2\lambda-\frac{\lambda^2}{2g}.
 \label{eqnB6}
\end{equation}
Choose $Y$ beyond all the minima in~\eqref{eqnB6}, with $U_\lambda(Y)>\sup K+1$ for every $|\lambda|\leq L$. 
By~\eqref{eq:quarticpotentialexp}, $W_{g,\lambda}(Y)>\sup K$ for sufficiently large $g$. 
Monotonicity beyond the minimum then places all allowed sets for $\xi\in K$ inside $[0,Y]$. Both the exact and quartic tails beyond $Y$ are forbidden.

Introduce
\begin{equation}
 Q_{h,\lambda}^{\rm bc}=-\frac{h^2}{2}\partial_y^2+U_\lambda(y),
 \label{eq:quarticoperator}
\end{equation}
with either Dirichlet or Neumann condition at $0$, as inherited from parity. Denote its strict count by $N_{h,\lambda}^{\rm bc}$. Imposing a boundary at $Y$ changes the count by at most one. Dirichlet--Neumann bracketing and \eqref{eq:quarticpotentialexp} give
\begin{equation}
 N_{h,\lambda}^{\rm bc}(\xi-Ch^2)-C_0
 \le N_{-,s}(e_{\rm sep}+\xi)
 \le N_{h,\lambda}^{\rm bc}(\xi+Ch^2)+C_0.
 \label{eq:quarticscalarbracket}
\end{equation}
The diagonal rotation term is already covered by \eqref{eq:quarticschur}, since $d=O(g^{-2})=O(h^4)$.

\subsection{A uniform oscillation estimate}
The required scalar estimate is
\begin{equation}
 \left|N_{h,\lambda}^{\rm bc}(\xi)-\frac{I_\lambda(\xi)}{\pi h}\right|\le C_{L,\mathcal K},
 \qquad 0<h\le1,\quad |\lambda|\le L,\quad \xi\in\mathcal K.
 \label{eq:uniformquarticweyl}
\end{equation}
Its uniformity at the coalescing critical points can be seen by an exact dilation. Set
\begin{equation}
 A=\max\{h^{2/3},|\lambda|,|\xi|^{1/2}\},\qquad
 y=A^{1/2}t,\qquad
 \mu=\lambda/A,\quad \epsilon=\xi/A^2,\quad \eta=h/A^{3/2}.
 \label{eq:quarticdilation}
\end{equation}
Then $0<\eta\le1$, $|\mu|,|\epsilon|\le1$, and
\begin{equation}
 Q_{h,\lambda}^{\rm bc}\simeq A^2Q_{\eta,\mu}^{\rm bc},\qquad
 I_\lambda(\xi)=A^{3/2}I_\mu(\epsilon).
 \label{eq:quarticexactdilation}
\end{equation}
If $A=h^{2/3}$, the effective parameter is $\eta=1$. The potentials are uniformly bounded below by $t^4/32-C$ and their allowed intervals are bounded. Both the count and the action are then $O(1)$. This includes the genuine quartic region
\begin{equation}
 |\lambda|\lesssim h^{2/3},\qquad |\xi|\lesssim h^{4/3},\qquad
 y=h^{1/3}t.
 \label{eq:quarticnormalform}
\end{equation}

In the remaining cases,
\begin{equation}
 \max\{|\mu|,|\epsilon|^{1/2}\}=1.
 \label{eq:quarticcompactboundary}
\end{equation}
This compact parameter set excludes the degenerate point $(\mu,\epsilon)=(0,0)$. Parameters $\eta$ bounded away from zero are controlled by the same confining comparison, so it suffices to consider small $\eta$. If $\epsilon=-1$, the energy lies below every potential minimum and the count and action vanish. If $\epsilon=1$, all turning points are simple with uniformly nonzero slope. 
On the normalized parameter set~\eqref{eq:quarticcompactboundary}, the remaining critical configurations are
\begin{equation}
 (\mu,\epsilon)=(1,0),\qquad (1,-1/4),\qquad (-1,0).
 \label{eq:quarticcriticalset}
\end{equation}
They are, respectively, the boundary saddle, the interior minimum and the boundary minimum. The corresponding curvatures of $U_\mu$ are $-1/2$, $1$ and $1/2$.

Here is a direct oscillation bound near these configurations. Put
$p(t)=\sqrt{2(\epsilon-U_\mu(t))}$ on an allowed interval. With $z=\eta^{-1}\int p(t)\,\dd t$ and $w=p^{1/2}u$, the scalar eigenvalue equation becomes
\begin{equation}
 \frac{\dd^2w}{\dd z^2}
 +\left[1+\eta^2\left(\frac{3(p')^2}{4p^4}-\frac{p''}{2p^3}\right)\right]w=0.
 \label{eq:quarticliouville}
\end{equation}
The difference between its Pr\"ufer phase and $z$, and hence the error in counting zeros apart from an endpoint constant, is bounded by
\begin{equation}
 C\eta\int\left(\frac{|p''|}{p^2}+\frac{(p')^2}{p^3}\right)\dd t.
 \label{eq:quarticphaseerror}
\end{equation}
At a simple turning point with slope magnitude $a_0$, remove a layer of width
$\ell_0=(\eta^2/a_0)^{1/3}$. Since $p\asymp(a_0\ell)^{1/2}$ there, the contribution outside the layer is bounded by
\begin{equation}
 C\eta a_0^{-1/2}\int_{\ell_0}\ell^{-5/2}\,\dd\ell=O(1).
 \label{eq:quarticairyerror}
\end{equation}
The removed layer has action $O(\eta)$ and contains $O(1)$ zeros by Sturm comparison.

When the energy lies within $C\eta$ of one of the critical values in \eqref{eq:quarticcriticalset}, remove a core of width $C_1\sqrt\eta$ about the critical point. In the core, $p^2\le C_2\eta$, so its action divided by $\eta$ and its zero count are both bounded. At a minimum this core contains the entire nearby allowed set. At the saddle, the exterior has $p\asymp t$ near the origin, and its contribution to \eqref{eq:quarticphaseerror} is bounded by
\begin{equation}
 C\eta\int_{C_1\sqrt\eta}^{c}t^{-3}\,\dd t=O(1).
 \label{eq:quarticsaddleerror}
\end{equation}
The outer turning point remains simple and is treated by \eqref{eq:quarticairyerror}.

For energy distance $d$ satisfying $C\eta<d<c$ from a critical value, any local turning points have distance from the critical point and slope magnitude of order $\sqrt d$. Their removed layers have width $\ell_0\asymp\eta^{2/3}d^{-1/6}$, with $\ell_0/\sqrt d\asymp(\eta/d)^{2/3}$. Choose $C$ large enough that these layers lie inside disjoint neighborhoods of width a fixed small multiple of $\sqrt d$. Within each such neighborhood, \eqref{eq:quarticairyerror} gives an $O(1)$ phase error outside the layer.

On the remaining part of a critical neighborhood, put $t=t_c+\sqrt d\,v$ and $p(t)=\sqrt d\,\widetilde p_d(v)$. The phase-error bound becomes
\begin{equation}
 \frac{C\eta}{d}\int
 \left(\frac{|\widetilde p_d''|}{\widetilde p_d^2}
       +\frac{(\widetilde p_d')^2}{\widetilde p_d^3}\right)\dd v.
 \label{eq:quartictransitionerror}
\end{equation}
After the turning-point neighborhoods have been removed, the integral is uniformly bounded. At a minimum the remaining scaled interval is bounded. At the saddle the integrand decays as $O(|v|^{-3})$ along the allowed exterior. Above the saddle the same estimate holds without an inner real turning point. Below a minimum both the count and action vanish. The regular region outside these critical neighborhoods is treated by the simple-turning-point estimate. This covers the transition from the critical cores to the regular region with one constant.

The number of cuts is uniformly bounded, since the quartic potential has at most two positive turning points. Each cut changes a scalar count by at most one, and forbidden intervals are disconjugate. Sturm oscillation and \eqref{eq:quarticexactdilation} now give \eqref{eq:uniformquarticweyl}. In the original variables the quadratic cores have spatial scale $h^{1/2}|\lambda|^{-1/4}$ and energy scale $h\sqrt{|\lambda|}$. At $|\lambda|\asymp h^{2/3}$ these become $h^{1/3}$ and $h^{4/3}$, exactly the quartic scales in \eqref{eq:quarticnormalform}.

\subsection{Returning to the Rabi count}
It remains to control the $O(h^2)$ energy displacements in \eqref{eq:quarticscalarbracket}. Uniformly on bounded parameter and energy sets,
\begin{equation}
 |I_\lambda(\xi+\zeta)-I_\lambda(\xi-\zeta)|\le C\zeta^{3/4},
 \qquad 0<\zeta<1.
 \label{eq:quarticholder}
\end{equation}
To see this without losing uniformity as roots merge, put $P(y)=\xi-U_\lambda(y)$. Factorization over its four complex roots gives
\begin{equation}
 |\{y\in[0,Y]:|P(y)|\le u\}|\le C u^{1/4}.
 \label{eq:quarticsublevel}
\end{equation}
Indeed, if the product of the four root distances is at most $16u$, one distance is at most $(16u)^{1/4}$. The real set is contained in four intervals of that length up to a fixed factor. In $|P|\le2\zeta$, the change in the square root is $O(\zeta^{1/2})$, giving $O(\zeta^{3/4})$. On the positive dyadic shells $2^k\zeta< P\le2^{k+1}\zeta$, the contribution is bounded by
\begin{equation}
 C\zeta(2^k\zeta)^{-1/2}(2^{k+1}\zeta)^{1/4}
 \le C'\zeta^{3/4}2^{-k/4}.
\end{equation}
Summation proves \eqref{eq:quarticholder}. The exponent is attained at the quartic minimum.

An $O(h^2)$ threshold shift therefore changes $I_\lambda/(\pi h)$ by $O(h^{1/2})$. Combining \eqref{eq:quarticschur}, \eqref{eq:quarticscalarbracket} and \eqref{eq:uniformquarticweyl} gives
\begin{equation}
 \N_s(e_{\rm sep}+\xi)=\frac{I_\lambda(\xi)}{\pi h}+O(1)
 =\frac{\sqrt g}{\pi}I_\lambda(\xi)+O(1).
 \label{eq:quarticcumulative}
\end{equation}
Subtraction at $a$ and $b$ gives \eqref{eq:quarticcount}.

\section{Numerical methods and validation}\label{app:numerics}
The numerical spectra are computed from leading $M\times M$ principal sections of the parity Jacobi matrices \eqref{eq:jacobi}. Each reported count is compared at two truncation dimensions. Agreement is a convergence check and does not itself bound the error from the omitted infinite tail. For the four-level example \eqref{eq:fourlevel}, outward-rounded arithmetic and an explicit tail estimate additionally validate the infinite-operator counts.

\subsection{Interval counts and the four-level example}\label{app:numerics-fixed}
For the counts in $[0,1)$ we use $g=2^k$ for $k=2,\ldots,10$, $\Delta=g^2-1/2$ and
\begin{equation}
 M_1=6g^2+400,\qquad M_2=8g^2+600.
 \label{eq:fixedbincutoffs}
\end{equation}
Selected eigenvalues are computed by tridiagonal bisection using SciPy's \texttt{eigh\_tridiagonal} with LAPACK driver \texttt{stebz}. We select the enlarged shifted interval $[-10^{-5},1+10^{-5}]$ and then impose $0\le x<1$. No selected level lies within $10^{-7}$ of $x=0$ or $x=1$. The two dimensions give identical interval counts. For \eqref{eq:fourlevel}, the dimensions are $M_1=6291856$ and $M_2=8389208$, and the eigenvalues agree to the displayed precision. An independently implemented long-double Sturm recurrence reproduces these counts. We strengthen this comparison as follows.

Fix $g=1024$, $\Delta=g^2-1/2$ and $M=M_2$. For $x\in[0,1]$, let $T_s^{(M)}(x)$ be the compression of $J_s+g^2-x$ to oscillator indices $m\ge M$. Bounding the off-diagonal row sums gives
\begin{equation}
 T_s^{(M)}(x)\ge\eta_M(x)I,
 \qquad \eta_M(x)=\frac M2-2g^2-x>0.
 \label{eq:numericaltailgap}
\end{equation}
Indeed, $g(\sqrt m+\sqrt{m+1})\le(m+1)/2+2g^2$, while the diagonal is at least $m+1/2-x$. Eliminating the positive tail changes only the last diagonal entry of the finite head, decreasing it by an amount between zero and $g^2M/\eta_M(x)$.

The finite-head negative inertia is obtained from the Sturm pivots
\begin{equation}
 p_0=g^2+s\Delta-x,
 \qquad
 p_m=m+g^2+s(-1)^m\Delta-x-\frac{g^2m}{p_{m-1}}.
 \label{eq:validatedsturm}
\end{equation}
We enclose every pivot using outward-rounded arithmetic with both 64-bit and 113-bit significands. The coefficients and tested thresholds are exact dyadic numbers, and the squared off-diagonal entries $g^2m$ are exact integers. No pivot enclosure contains zero. At $x=0$ and $1$, for either parity, the final pivot remains positive after subtracting the largest allowed tail correction. Across these endpoint calculations, the corrected final-pivot lower bound exceeds $3.2497\times10^6$. In the 64-bit runs, the minimum distance of a pivot enclosure from zero, divided by its width, exceeds $2.4\times10^5$. Thus the omitted tail cannot change the negative inertia, and the finite-head counts equal the infinite-operator counts. Both precisions give the counts in Table~\ref{tab:validatedcounts}.

\renewcommand{\thetable}{C.\arabic{table}}
\setcounter{table}{0}

\begin{table}[htbp]
\centering
\label{tab:validatedcounts}
\begin{tabular}{crrr}
\toprule
Parity & Levels with $x<0$ & Levels with $x<1$ & $B_0^s$\\
\midrule
$+$ & $333\,772$ & $333\,775$ & $3$\\
$-$ & $333\,772$ & $333\,776$ & $4$\\
\bottomrule
\end{tabular}
\caption{Certified cumulative counts at $g=1024$ and $\Delta=1048575.5$. Both arithmetic precisions give the same counts for the infinite Jacobi operator.}
\end{table}

The same tail argument and 113-bit pivot enclosures locate each of the four levels in \eqref{eq:fourlevel} in a disjoint interval of width less than $2.1\times10^{-8}$. The lowest lies above $0.00494057$, safely away from the integer endpoint. These checks validate this particular four-level example for the infinite Jacobi operator. The other numerical examples retain the stated finite-section convergence checks.

\subsection{A finite-coupling counterexample}
For a direct connection with a reported counterexample, take $g=6$ and $\Delta=30$ \cite{Lanuza2026}. Then $\beta=5/6$, $c=0$ and $x_{\rm sep}=5.5$. The negative-parity shifted levels in $[4,6)$ are
\begin{equation}
 4.01069247,\quad4.73993239,\quad5.36451580,\quad5.94617587.
 \label{eq:counterexamplelevels}
\end{equation}
Thus $B_4^-+B_5^-=4$ in a window containing the separatrix. These four levels are checked by tridiagonal bisection at dimensions $M=350$ and $550$. The sufficient tail boundary at these parameters is $n=66$. This is a finite-section check of the reported example.

\subsection{Macroscopic density}\label{app:numerics-density}
For the numerical evaluation, a squared-sine substitution removes the simple turning-point singularities and a 160-point Gauss--Legendre rule evaluates the resulting integral. Above the upper-band onset we use the exact value \eqref{eq:rhoone}.

\subsection{Quartic crossover}\label{app:numerics-quartic}
The finite-section counts for \eqref{eq:quarticexample} are unchanged on increasing the Jacobi dimension from $76\,560$ to $114\,840$. The same two dimensions are used for Figure~\ref{fig:quartic}. These are truncation-convergence checks, without the infinite-operator certification used for the four-level example in $[0,1)$.

For Figure~\ref{fig:quartic} we take $g=64$, $\lambda=-16,-12,\ldots,16$ and $[a,b)=[0,16)$. The example \eqref{eq:quarticexample} uses $\lambda=2$ and $[a,b)=[-2,2)$. The strict finite-section counts are obtained by the Sturm recurrence
\begin{equation}
 z_0=s\Delta-e,\qquad
 z_m=m+s\Delta(-1)^m-e-\frac{g^2m}{z_{m-1}},
 \qquad 1\le m<M.
 \label{eq:sturmrecurrence}
\end{equation}
The number of negative pivots gives the count below $e$, away from finite-section eigenvalues. Pivots smaller than $10^{-200}$ in magnitude are replaced by $-10^{-200}$ to continue the recurrence. Subtracting the endpoint counts gives the displayed window counts, which agree at both dimensions.

The quartic action \eqref{eq:Fquartic} is evaluated over its classically allowed interval on the half-line. We use its exact value at $\xi=0$ and adaptive quadrature otherwise, with absolute and relative tolerances $2\times10^{-10}$.

\FloatBarrier

\end{document}